\documentclass[11pt,twoside]{article}
\usepackage{asp2004}
\usepackage{psfig}
\usepackage{epsf}
\usepackage{graphics}
\usepackage{lscape}
\begin{document}
\title{A New Kind of Nova}
\author{J. L. Sokoloski, S. J. Kenyon, \& A. K. H. Kong}
\affil{Harvard-Smithsonian CfA, 60 Garden St., Cambridge, MA 02138, USA}
\author{B. R. Espey}
\affil{Physics Department, Trinity College, Dublin 2, Ireland}
\author{S. R. McCandliss}
\affil{Dept. of Physics \& Astronomy, JHU,
3400 North Charles St., Baltimore, MD 21218, USA}
\author{C. D. Keyes}
\affil{STScI, 3700 San Martin Dr., Baltimore, MD 21218, USA}
\author{W. Li \& A. V. Filippenko}
\affil{Astronomy Dept., 601 Campbell Hall, U. C. Berkeley, Berkeley,
CA 994720, USA}
\author{J. Aufdenberg}
\affil{NOAO, 950 North Cherry Ave., Tucson, AZ 85719, USA}
\author{C. Brocksopp}
\affil{MSSL, University College London, Dorking, Surrey, RH5 6NT, UK}
\author{C. R. Kaiser \& P. A. Charles}
\affil{Dept. of Physics and Astro., Southampton University, 
SO17 1BJ, UK}
\author{R. P. S. Stone}
\affil{UCO/Lick Observatory, U. C. Santa Cruz, Santa Cruz, CA
95064, USA}

\vspace{-0.5mm}
\begin{abstract}
We performed extensive, multi-wavelength observations of the
prototypical symbiotic star Z Andromedae between 2000 and 2003, during
a large eruption.  The rise to optical maximum occurred in three
distinct stages.  During the first stage, the rise was very similar to
an earlier, small outburst which we determined was due to an accretion-disk
instability. In the second stage, an optically thick shell of material
was ejected, and in the third stage, the shell cleared to reveal a
white dwarf whose luminosity was roughly $10^4 L_{\sun}$.  We suggest
that the outburst was powered by an increase in the rate of nuclear
burning on the white-dwarf surface, triggered by a sudden burst of
accretion.  This outburst thus combined elements of both dwarf novae
and classical novae.
\end{abstract}
\thispagestyle{plain}

\section{Introduction}

Determining the cause of classical symbiotic-star outbursts has been a
long-standing challenge.  These eruptions recur too frequently for
nova-like thermonuclear runaways, and their peak luminosities appear
to be too 
large for dwarf-nova-like disk instabilities.
Although quasi-steady nuclear burning is probably present on the
surface of the white dwarfs (WDs) in most symbiotics, classical
symbiotic outbursts are also distinctly different from the long-term
variability seen in supersoft X-ray sources.

Symbiotic stars are interacting binaries in which material is
transferred from an evolved red-giant star to a more compact, hot
star, usually a WD \citep[see, e.g.,][]{kbook,lapalmabook}.  In most
symbiotics, the red giant under-fills its Roche lobe, and the mass
transfer proceeds via gravitational capture of the red giant's wind.
An accretion disk may or may not form
\citep{livio88}.  Radiation from the accreting WD partially
ionizes the nebula formed by the red-giant wind.  This ionized nebula
gives rise to intense emission lines.  Because typical quiescent-state
WD luminosities in symbiotic stars are roughly $10^3\, L_{\sun}$
\citep[e.g.,][]{mur91}, quasi-steady nuclear shell burning 
is thought to be taking place on the WD surface in the majority of systems
\citep[see
also][]{vdh92,sbh01}.

There are at least three types of symbiotic-star outbursts.  Slow
novae and recurrent novae appear to be due to thermonuclear runaways
on the WD surface.  The nature of the more common classical symbiotic
outbursts, however, is not yet known.  
These eruptions can recur as frequently as every few years, and can
brighten in the optical by one to a few magnitudes. 
\cite{kbook} found that dwarf-nova-like accretion-disk instabilities
in a symbiotic star can only produce outbursts of up to about one
magnitude, so outburst modeling efforts have generally focused on
expansion of the WD photosphere at constant bolometric luminosity (due
to the accretion rate rising above the maximum value for
steady-burning), or hydrogen shell flashes.
The amount of mass lost during classical symbiotic outbursts has
implications for whether symbiotic WDs can accrete enough material to
explode as Type Ia supernovae.  Since the mass loss can take the form
of a collimated jet, the nature of classical symbiotic outbursts is
also linked to the issue of jet formation.

\section{Multi-wavelength Monitoring of Z Andromedae}

The hot components (e.g., white-dwarf plus accretion disk) in
symbiotic stars emit the bulk of their energy in the FUV.
However, they also radiate significantly at radio through X-ray
wavelengths, and important diagnostics are found in each of these
observational regimes.  To investigate the nature and cause of
classical symbiotic-star outbursts, we performed multi-wavelength
observations, including observations with the $FUSE$ (9 spectra from
2000 Nov 16 to 2003 Aug 4), $Chandra$ (1 observation on 2000 Nov 13),
and $XMM$ (2 observations on 2001 Jan 29 and Jun 11) satellites, the
VLA (9 observations between 2000 Oct 13 and 2003 Jul 24) and MERLIN (6
observations between 2001 Jan 28 and 2002 May 6) radio
interferometers, and extensive ground based optical spectroscopy and
photometry, during the recent 2000-2003 activity phase of Z And.
Additional optical spectroscopic coverage extended back to
1994. The full set of observations is described in detail in
\cite{soko05}.

\section{Results}

\subsection{I. Hot-Component Effective Temperature}

\begin{figure}[!ht]
\plotone{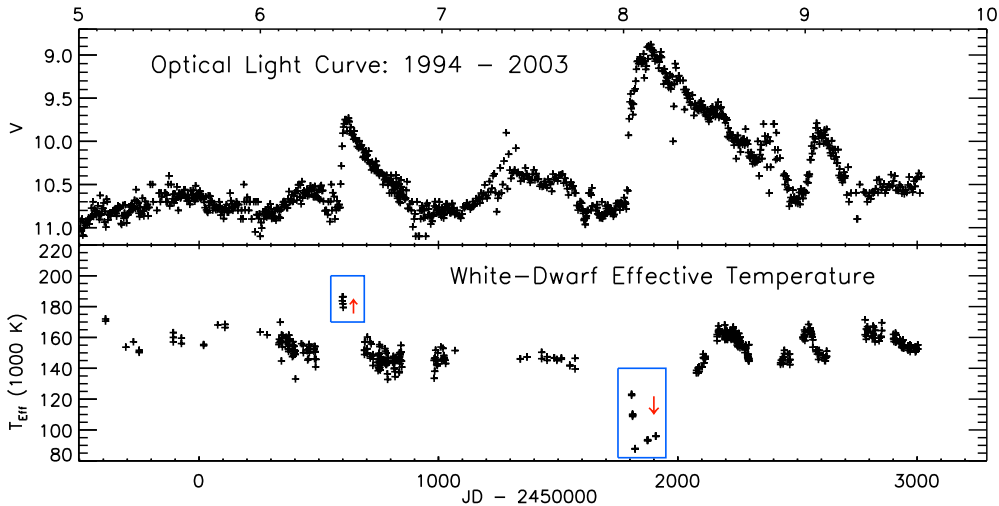}
\caption{Top: Long-term optical light curve of Z And, from the American
Association of Variable Star Observers (AAVSO).  The light curve shows
the small outburst in 1997 and the larger 2000-2002 event (and
subsequent rebrightening).  Bottom: Estimates of the hot-component
effective temperature from the ratio of He II 4686 and H$\beta$
nebular emission lines. Orbital phase from the ephemeris of
\citet{mk96} is shown on the top. \label{fig:thot}}
\end{figure}

In the top panel of Fig.~\ref{fig:thot}, we show the optical light
curve of Z And from 1994 to 2003, from the American Association of
Variable Star Observers (AAVSO).  In the bottom panel, we plot
estimates of the hot-component effective temperature, $T_{hot}$, that
we derived from 274 nights of optical spectroscopy.  It is clear from
the disparate behavior of $T_{hot}$ during the small outburst in 1997
and the larger eruption in 2000-2002 that these two events were quite
different.  Therefore, our first conclusion is that not all classical
symbiotic star outbursts are due to the same physical mechanism.

We estimated the effective temperature of the source of ionizing
photons (the hot component)
from 1994 to 2003 using a method based on that of \cite{iijima81} plus
examination of the highest ionization-potential species present in the
optical and contemporaneous FUV spectra \citep[for details,
see][]{soko05}.  The behavior of the first outburst covered by our
optical spectral monitoring is consistent with an accretion-disk
instability, or dwarf nova.  The evidence for this interpretation
includes: the shape and size of the optical outburst (fast rise plus
exponential decay, and $\Delta V \le 1$); the evolution of the
hot-component effective temperature; the roughly constant fractional
amplitude of the photometric oscillation at the WD spin period due to
magnetic accretion (after subtraction of the contribution from the red
giant) throughout the outburst \citep{sb99}; and the expectation that
the potentially large disks in symbiotic stars should in fact be
unstable.

In the 2000-2002 event, the rise to optical maximum proceeded in three
distinct stages, the optical brightness increased by well over the one
magnitude that can be produced by a disk instability in a symbiotic
\citep{kbook}, and the hot-component effective temperature evolved in a
rather complex way.  The 2000-2002 eruption was therefore not
a simple dwarf nova.

\subsection{II. Three-Stage Rise}

\begin{figure}[!ht]
\plotone{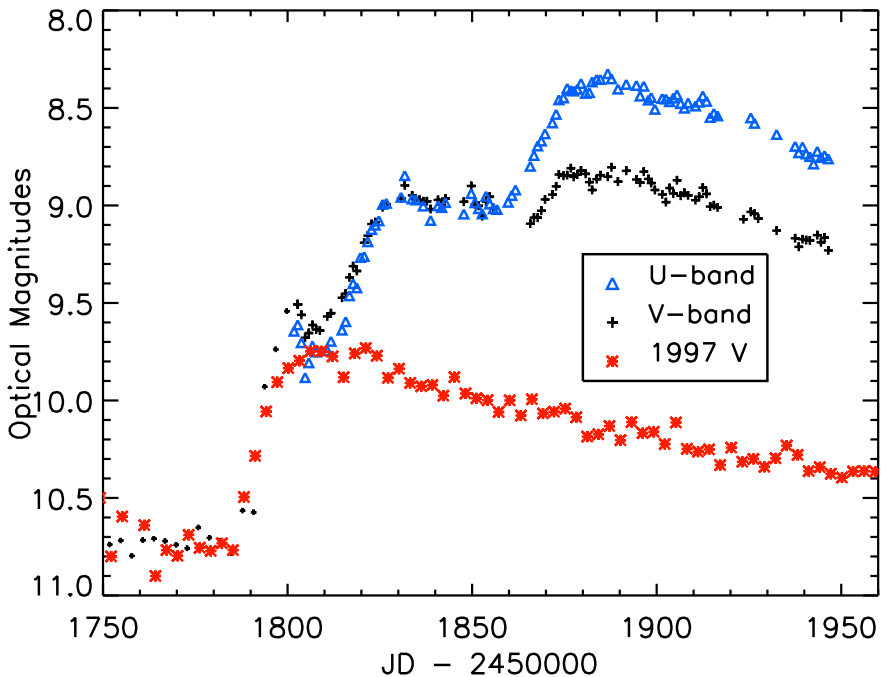}
\caption{$U$ 
and $V$-band light curves from the early part of the 2000-2002
outburst of Z And, with the 1997 $V$-band light curve shifted in time
and over-plotted.  The optical rise initially follows the same course
for the two events.  In 2000, however, the first-stage rise proceeds
beyond the maximum level reached in the 1997 event, and the outburst
evolution subsequently takes a different path.  The triangles and
crosses are our data from the Katzman Automatic Imaging Telescope
\citep[KAIT;][]{wli2000,filip2001}, and the dots and $\ast$ symbols
are data from the AAVSO.
\label{fig:oplot}}
\vspace{-0.4cm}
\end{figure}

The 2000-2002 eruption of Z And began like the
smaller event in 1997.  Fig.~\ref{fig:oplot}, in which the 1997 and
2000 light curves
are overlayed, shows the similarity between the first stage of the
2000-2002 event and the rise to maximum in 1997.  This initial
correspondence suggests that the physical trigger mechanism was the
same for both the large 2000-2002 outburst and the small 1997
eruption.  
At the very beginning of the 2000-2002 event, the
$U$-band flux rose quickly and the $U-B$ color became more blue
\cite[see][for more complete coverage of this blue
spike]{sko02}.  Since the $U$ brightness tends to be dominated by
reprocessed ionizing radiation from the hot component, the 2000-2002
outburst probably began with a process that rapidly increased the
luminosity of the hot component.  The blue flare early in the
2000-2002 event provides a glimpse of an initial $T_{hot}$ increase in
2000 (like the $T_{hot}$ increase seen in 1997) before the main
effective-temperature dip later in the 2000-2002 eruption.  We
therefore conclude that the 2000-2002 outburst began with a disk
instability, as in 1997.
 
During the second stage of the 2000-2002 outburst, $T_{hot}$ dropped
(see Fig.~\ref{fig:thot}) and the $U-B$ color reddened, first sharply,
and then more gently.  The first $FUSE$ spectrum, taken near the end
of the second stage of the outburst, was dominated by blue-shifted
absorption profiles indicating significant outflow
\citep[see][]{soko02}, and the ratio of P V line components was as
expected for an optically thick plasma.  The uncharacteristically low
5 GHz radio flux density of 0.42 mJy measured during the second stage
of the outburst (on 2000 Oct 13) suggests that the ionized,
bremsstrahlung-emitting nebula was smaller than usual due to a reduced
flux of ionizing photons from the WD.  In addition, on 2000 Nov 23 (at
the beginning of the third stage of the outburst), the optical
oscillation at the WD spin period was not detected.  Thus, the
magnetic hot spots on the surface of the WD appear to have been
hidden.  All of these phenomena can be explained by an ejection from
the surface of the WD of an optically thick shell of material, or
possibly the presence of an optically thick wind.

During the third stage of the rise to optical maximum in 2000,
$T_{hot}$ slowly began to increase again and the $U-B$ color became
more blue, the FUV spectrum moved from absorption to emission, the
radio flux rose, and the final increase in optical flux became
dominated by $U$-band emission.  These changes are consistent with the
clearing of material that had shrouded the FUV continuum in the second
stage of the outburst, to reveal a hot, luminous white dwarf.

\subsection{III. White-Dwarf Luminosity}

We find that the increase in WD luminosity during the 2000-2002
outburst of Z And was too large to have been produced by accretion
alone.  The most likely explanation is that the rate of
nuclear shell burning on the surface of the WD increased, probably as
the result of the sudden influx of fresh fuel.

To determine the bolometric luminosity of the WD throughout the
2000-2003 activity period, we estimated the WD radius, $R_{WD}$, at
the time of each of the 9 $FUSE$ observations by scaling the WD
photosphere models of \citet{barman00} to the extinction-corrected
$FUSE$ fluxes, using the effective temperatures described above, and
assuming a distance of 1 kpc and mass of 0.65 $M_{\sun}$
\citep{schsch97}.  We then obtained the WD luminosities using the
standard relation $L_{hot}=4\pi R_{WD}^2 \sigma T^4$.  The derived
hot-component luminosities are close to $10^4\, L_{\sun}$ from the end
of 2000 to late 2001.  For comparison, the Eddington luminosity for a
$0.65 M_{\sun}$ white dwarf is $L_{Edd} = 3 \times 10^4\, L_{\sun}\,
(M/0.65\, M_{\sun})$.

If the 2000-2002 outburst was entirely accretion powered, the required
accretion rate to produce $10 ^4\, L_{\sun}$ would be
$\dot{M} 
\approx 5 \times 10^{-5} M_{\sun}\, {\rm yr}^{-1}$,
where we adopt $R_{WD} = 0.1\, R_{\sun}$ as indicated by scaling the
FUV fluxes to photospheric models, and $M_{WD} = 0.65\, M_{\sun}$.  Such a high accretion rate would be
difficult to sustain in Z And for a full year.   If the outburst was nuclear
powered, on the other hand, only a few times $10^{-7}\, M_{\sun}$ of
fuel would be required to produce $L_{hot} \sim 10^4\, L_{\sun}$ for
one year, which is more reasonable.  If we take into account the kinetic
energy of the ejected mass
\citep{tomov03,brock04}, thermonuclear involvement in the outburst is
even more strongly indicated\footnote{\citet{kilpio} also suggest that 
nuclear shell burning played a role in the 2000-2002 outburst of Z
And, although their trigger mechanism is a collapse of the accretion
disk when the red-giant wind speed drops below a critical level.}.

\section{Combination Nova}

We see evidence for phenomena that usually occur in two distinct types
of cataclysmic variable star outbursts in the same symbiotic star
event.  The 2000-2002 outburst of Z And appears to have been triggered
by a disk instability, as in dwarf novae, and then powered by an
increase in nuclear shell burning on the WD, in a milder version of
the phenomenon that powers classical novae.
The outburst in Z And therefore combines elements of dwarf novae and
classical novae, and we refer to this new type of event as a
combination nova.

\acknowledgments

J. L. S. is supported by an NSF Astronomy and Astrophysics Postdoctoral
Fellowship under award AST 03-02055.


\begin{thebibliography}{}
\bibitem[Barman et al.(2000)]{barman00}Barman, T. S., Hauschildt,
P. H., Short, C., I., \& Baron, E. 2000, ApJ, 537, 946
\bibitem[Brocksopp et al.(2004)]{brock04}Brocksopp, C., Sokoloski,
J. L., Kaiser, C., Richards, A. M., Muxlow, T. W.  B., Seymour,
N. 2004, MNRAS, 347, 430
\bibitem[Corradi, Miko{\l}ajewska, \& Mahoney(2003)]{lapalmabook}
Corradi, R. L. M., Miko{\l}ajewska, J., \& Mahoney, T. J., eds., 2003,
Symbiotic Stars Probing Stellar Evolution, ASP Conf. Proceedings,
Vol. 303
\bibitem[Filippenko et al.(2001)]{filip2001}Filippenko, A. V., et al. 2001,
in {\it Small Telescope Astronomy on Global Scales}, eds. B. Paczynski
et al. (San Francisco: ASP), p121
\bibitem[Iijima(1981)]{iijima81}Iijima, T. 1981, in Photometric and
Spectroscopic Binary Systems, Proceedings of the NATO advanced Study
Institute (NATO ASI series, v. 69), p. 517
\bibitem[Kenyon(1986)]{kbook}Kenyon, S. J. 1986, ``The Symbiotic
Stars'', Cambridge University Press, Cambridge
\bibitem[Kilpio et al.(2005)]{kilpio}Kilpio, E., Bisikalo, D. V.,
Boyarchuk, A. a., \& Kuznetsov, O. A. 2005, this volume
\bibitem[Li et al.(2000)]{wli2000}Li, W., et al. 2000, in {\it Cosmic
Explosions}, 
eds. S. S. Holt \& W. W. Zhang (New York: American Institute of
Physics), p103
\bibitem[Livio(1988)]{livio88}Livio, M. 1988, in The Symbiotic
Phenomenon, Proc. of IAU Colloq. 103, J. Miko{\l}ajewska,
M. Friedjung, S. J. Kenyon, \& R. Viotti, eds.,  
p. 149
\bibitem[Miko{\l}ajewska \& Kenyon(1996)]{mk96}Miko{\l}ajewska, J., \&
Kenyon, S. J. 1996, AJ, 112, 1659
\bibitem[M{\"u}rset et al.(1991)]{mur91}M{\"u}rset, U., Nussbaumer,
H., Schmid, H. M., \& Vogel, M. 1991, A\&A, 248, 458
\bibitem[Schmid \& Schild(1997)]{schsch97}Schmid, H. M., \& Schild,
H. 1997, A\&A, 327, 219
\bibitem[Skopal et al.(2002)]{sko02}Skopal, A., V\v{a}nko, M.,
Pribulla, T., Wolf, M., Semkov, E., \& Jones, A. 2002, 
Contributions of the Astronomical Observatory Skalnate Pleso, vol. 32, 
no. 1, p 62
\bibitem[Sokoloski et al.(2005)]{soko05}Sokoloski, J. L., et al. 2005, 
in preparation
\bibitem[Sokoloski et al.(2002)]{soko02}Sokoloski, J. L. et al. 2002,
in The Physics of Cataclysmic Variables and Related Objects, ASP
Conf. Ser., 261, B. T. G{\"a}nsicke, K. Beuermann, \&
K. Reinsch, eds., p. 667
\bibitem[Sokoloski \& Bildsten(1999)]{sb99}Sokoloski, J. L., \&
Bildsten, L. 1999, ApJ, 517, 919
\bibitem[Sokoloski et al.(2001)]{sbh01}Sokoloski, J. L., Bildsten, L., 
\& Ho, W. C. G. 2001, MNRAS, 326, 553
\bibitem[Tomov et al.(2003)]{tomov03}Tomov, N. A., Taranova, O. G.,
Tomova, M. T. 2003, A\&A, 401, 669
\bibitem[van den Heuvel et al.(1992)]{vdh92}van den Heuvel, e. P. J.,
Bhattacharya, D., Nomoto, K., \& Rappaport, S. A. 1992, A\&A, 262, 97
\end{thebibliography}
\end{document}